\documentclass[a4paper]{article}
\usepackage{amsmath}
\usepackage{graphicx}
\usepackage{geometry}
\usepackage{floatrow}
\usepackage{layout}
\usepackage{amssymb}
\usepackage{multirow}
\usepackage{caption}
\usepackage{authblk}
 \usepackage{indentfirst}
\usepackage[mathscr]{euscript}
\usepackage{mathtools}
\usepackage{hyperref}
\usepackage{cleveref}
\nonstopmode
\begin{document}
\title{ Relation Between the Partial Derivatives of the Kinetic Energy in the Lagrangian and Hamiltonian Formalisms of Dynamics}
\author{Asghar Ali}

\affil{Department of Mathematical Sciences, Balochistan University of Information Technology,
Engineering and Management Sciences, Quetta 87300, Pakistan
Email:asghar.ali@buitms.edu.pk, asgharalikhannn@gmail.com}

\maketitle 

\begin{abstract}
The partial derivative of the kinetic energy of a dynamical system with respect to a generalized coordinate as it appears in the Lagrangian formalism is not equal to the derivative of the kinetic energy with respect to the same coordinate in the Hamiltonian formalism but differs by a sign. We find another exact relation between the two partial derivatives in the case of a conservative system. We also identify another form of kinetic energy whose partial derivative with respect to a generalized coordinate vanishes identically.\\
\emph{keywords:} Lagrangian, Hamiltonian, partial derivatives.
\end{abstract}

\section{Introduction}
Consider a system consisting of a single classical particle of mass $m$. The expressions for its kinetic energy in the Lagrangian formalism and Hamiltonian formalism are \cite{gol}
\begin{equation}\label{1}
T(r,\dot{r},\dot{\theta},\dot{z})=\frac{m}{2}(\dot{r}^{2}+r^{2} \dot{\theta}^{2} +\dot{z}^{2}), \qquad T(r,p_{r},p_{\theta},p_{z})=\frac{1}{2m}(p_{r}^{2}+\frac{p_{\theta}^{2}}{r^{2}} +p_{z}^{2}),
\end{equation}
where
$
p_{r}=m \dot{r}, \quad p_{\theta}=m r^{2} \dot{\theta}, \quad p_{z}=m \dot{z}
$
are the generalized momenta of the particle. Now, it follows from Eqs. \eqref{1} that

\begin{equation*}
mr\dot{\theta}^{2}=\frac{\partial}{\partial r}T(r,\dot{r},\dot{\theta},\dot{z}) \neq \frac{\partial}{\partial r} T(r,p_{r},p_{\theta},p_{z})=-mr\dot{\theta}^{2}.
\end{equation*}

It is confusing that $T(r,\dot{r},\dot{\theta},\dot{z})=T(r,p_{r},p_{\theta},p_{z})$ but
$$\frac{\partial}{\partial r} T(r,\dot{r},\dot{\theta},\dot{z}) \neq \frac{\partial}{\partial r} T(r,p_{r},p_{\theta},p_{z}).$$

What is amiss here? It is pointed out in \cite{gan} that the partial derivative of the kinetic energy of a dynamical system with respect to a generalized coordinate as it appears in the Lagrangian formalism is not equal to the derivative of the kinetic energy with respect to the same coordinate in the Hamiltonian formalism but differs by a sign. We encounter similar apparent inconsistencies in certain other coordinates systems. We find another form of kinetic energy which is not a function of the generalized coordinates for a conservative system in the next section.
\section{The Partial Derivatives of Kinetic Energy With Respect to a Generalized Coordinate}
Consider a conservative system with the Lagrangian and Hamiltonian,

\begin{equation*}
L(q, \dot{q(t)})=T(q(t), \dot{q}(t))-V(q(t)), \qquad H(q(t), p(t))=T(q(t), p(t))+V(q(t)),
\end{equation*}

where $q=(q^{1},...,q^{n})$, $\dot{q}=(\dot{q}^{1},...,\dot{q}^{n})$ and $p=(p_{1},...,p_{n})$ represent the generalized coordinates,  generalized velocities and generalized momenta of the system \cite{arn}. The Hamilton's equations for this system are

\begin{equation}\label{2}
\frac{\partial H}{\partial q^{j}}=-\dot{p}_{j}, \qquad \frac{\partial H}{\partial p_{j}}=\dot{q}^{j},
\end{equation}

where $j=1,...,n.$ The Lagrange's equations for this system are

\begin{equation}\label{3}
\frac{d}{dt}\left(\frac{\partial L}{\partial \dot{q}^{j}}\right)-\frac{\partial L}{\partial q^{j}}=0, \qquad j=1,...,n.
\end{equation}

Since $$p_{j}=\frac{\partial L}{\partial \dot{q}^{j}}$$ \cite{arn}, it immediately follows from Eqs.\eqref{3} that

\begin{equation} \label{4}
\frac{\partial L}{\partial q^{j}}=\dot{p}_{j}, \qquad j=1,...,n.
\end{equation}
Substituting Eqs.\eqref{4} into the first of Hamilton's equations in Eqs. \eqref{2}, we find after simplification that
\begin{equation}\label{5}
\frac{\partial }{\partial q^{j}}(T(q(t),\dot{q}(t))+T(q(t),p(t))=0 \quad {\text{or}} \quad \frac{\partial T(q(t), \dot{q}(t)) }{\partial q^{j}}=-\frac{\partial T(q(t), p(t)) }{\partial q^{j}},
\end{equation}
for any $j\in \{1,...,n\}.$ \\

It is very tempting to equate the two kinetic energy terms in the first of preceding equations and deduce that

\begin{equation}
\frac{\partial T }{\partial q^{j}}=0, \qquad j=1,...,n,\label{6}
\end{equation}
but this is not correct in general. Further, it would suggest that the kinetic energy of a conservative system can never be a function of the generalized coordinates which is obviously nonsensical. However, Eqs. \eqref{6} would follow when the kinetic energy of the system is not a function of the generalized coordinates in the relevant  coordinate system. \\

Moreover, the inequality of the two partial derivatives in Eqs. \eqref{5} follows from the obvious fact that the partial derivative in the left of these equations is evaluated at fixed generalized velocities and fixed generalized coordinates except $\dot{q}^{j}$, while partial derivative in the left of these equations is evaluated at fixed generalized momenta and fixed generalized coordinates other than $\dot{q}^{j}$. It also follows from the first of Eqs.\eqref{5} that  Kinetic energy can also be expressed in another form as

\begin{equation}\label{7}
M=\frac{1}{2}(T(q,\dot{q})+T(q,p)).
\end{equation}
Using Eqs. \eqref{5}, we get from Eq. \eqref{7}
\begin{equation}\label{8}
dM=\frac{1}{2}\left( \frac{\partial T(q,\dot{q}) }{\partial \dot{q}^{j}}d\dot{q}^{j}+\frac{\partial T (q,p)}{\partial p^{j}}dp^{j}\right),
\end{equation}
where summation over the repeated index $j$ is assumed.
Now, it follows from Eq. \eqref{8} that

\begin{equation*}
M=T(p,\dot{q}).
\end{equation*}

This equation shows that this form of the kinetic energy for a conservative system is independent of the generalized coordinates, and hence its partial derivative with respect to a generalized coordinate vanishes identically. Thus, Eq. \eqref{7} gives yet another form of kinetic energy, whose partial derivatives with respect to a generalized coordinate is always zero. \\

\section{Another Look at the Preceding Section}
Since the generalized momenta are functions of the generalized velocities and generalized coordinate, namely

\begin{equation}\label{9}
p=p(q, \dot{q})
\end{equation}

and the kinetic energy in the Lagrangian and Hamiltonian formalism has the same value, we have

\begin{equation}\label{10}
T(q^{1},...,q^{n},\dot{q}^{1},...,\dot{q}^{n})=T(q^{1},...,q^{n},p_{1},...,p_{n}).
\end{equation}

Now using Eq. \eqref{9}, we get from Eq. \eqref{10}

\begin{equation}\label{11}
\frac{\partial T(q^{1},...,q^{n},\dot{q}^{1},...,\dot{q}^{n})}{\partial q^{j}}=\frac{\partial T(q^{1},...,q^{n},p_{1},...,p_{n})}{\partial q^{j}}+\frac{\partial T(q^{1},...,q^{n},p_{1},...,p_{n})}{\partial p_{k}}\frac{\partial  p_{k}}{\partial q^{j}},
\end{equation}
where summation over the repeated index $k$ is assumed.
Applying this result to Eq. \eqref{1}, we get

\begin{equation*}
\frac{\partial}{\partial r}T(r,\dot{r},\dot{\theta},\dot{z})=\frac{\partial}{\partial r} T(r,p_{r},p_{\theta},p_{z})+\frac{\partial p_{\theta}}{\partial r}\frac{\partial}{\partial p_{\theta}} T(r,p_{r},p_{\theta},p_{z}),
\end{equation*}

which reduces to the identity, $mr\dot{\theta}^{2}=mr\dot{\theta}^{2}.$ \\

\section{Conclusion}
Lagrange's and Hamilton's equations show that the partial derivatives of the kinetic energy with respect to a generalized coordinates in the Lagrangian and Hamiltonian formalism are not equal but differ by a sign \cite{gan}.  We found another exact relation between the partial derivatives of the kinetic energy of a conservative system with respect to a generalized coordinate as it appears in the Lagrangian and Hamiltonian formalisms. This relation is given by  Eqs.\eqref{11}. We also found another form of kinetic energy that is a function of the generalized velocities and generalized momenta, and does not depend on the generalized coordinates. This form of kinetic energy is given by Eq. \eqref{7}.
\subsection*{Acknowledgement} I am very grateful to Nivaldo A. Lemos for sharing with me  \cite{gan} and his comment \cite{niv} on \cite{gan}. This helped me restructure this paper.

\end{document}